# Regular, beating and dilogarithmic breathers in biased photorefractive crystals

C.A. Betancur-Silvera[1], A. Espinosa-Cerón[2], B.A. Malomed[3,4] and J. Fujioka[1,*]

[1]Instituto de Física, Dpto. de Sistemas Complejos, Universidad Nacional Autónoma de México, Apdo. Postal 20-364, 01000 CDMX, México

[2]Facultad de Ciencias, Universidad Nacional Autónoma de México, 04510 CDMX, México

[3]Department of Physical Electronics, Faculty of Engineering, Tel Aviv University, Israel

[4]Instituto de Alta Investigación, Universidad de Tarapacá, Casilla 7D, Arica, Chile

[*]Corresponding author: fujioka@fisica.unam.mx

**Abstract**

The propagation of light beams in photovoltaic pyroelectric photorefractive crystals is modelled by a specific generalization of the nonlinear Schrödinger equation (GNLSE). We use the variational approximation (VA) to predict the propagation of solitary-wave inputs in the crystal, finding that the VA equations involve the dilogarithm special function. The VA predicts that solitons and breathers exist, and the Vakhitov-Kolokolov criterion predicts that the solitons are stable solutions. Direct simulations of the underlying GNLSE corroborates the existence of such stable modes. The numerical solutions produce both regular breathers and ones featuring *beats* (long-period modulations of fast oscillations). In the latter case, the Fourier transform of amplitude oscillations reveals a nearly discrete spectrum characterizing the beats dynamics. Numerical solutions of another type demonstrate spontaneous splitting of the input pulse in two or several secondary ones.

## 1. Introduction

The propagation of light in photorefractive materials has been widely studied since they were discovered by Ashkin in 1966 [1]. Among other topics addressed in this context is the propagation of self-trapped light beams in the form of spatial solitons, whose transverse width remains bounded. In particular, they are identified as breathers if the beam´s width and amplitude oscillate in the course of the propagation. It is known that photorefractive crystals support steady-state solitons in different settings, such as:

a) biased nonphotovoltaic photorefractive crystals (*screening solitons*) [2-6];

b) unbiased photovoltaic photorefractive crystals (*photovoltaic solitons*) [3-9];

c) biased photovoltaic photorefractive crystals (*screening photovoltaic solitons*) [4-6, 10];

d) unbiased photovoltaic photorefractive crystals affected by temperature variations (*photovoltaic pyroelectric solitons*) [5];

e) biased photovoltaic photorefractive crystals affected by temperature variations (*screening photovoltaic pyroelectric solitons*) [6];

f) photovoltaic crystals exhibiting two-photon photorefractive effect [11].

The propagation of breathers in photorefractive crystals has received less attention, in comparison to breathers which appear in other systems [12-16]). In the present work, we focus on the breather dynamics.

The propagation of breathers and solitons in photorefractive crystals, apart from their possible applications [17], is particularly interesting from a theoretical point of view, as it is modelled by equations similar to the nonlinear Schrödinger (NLS) equation. In the present work we study the propagation of light beams in a biased photovoltaic photorefractive crystal featuring the pyroelectric effect. It is known that this medium maintains spatial solitons, which are governed by a generalized NLS equation (GNLSE), that includes the usual diffractive term (the second derivative with respect to the transversal coordinate), Kerr-type and saturable nonlinearities, produced by the pyroelectric and photovoltaic effects, respectively, and a nonlinear term induced by the external field [18]. In this work we introduce a *saturation parameter*, which is a measure of the combined effect of the three distinct nonlinearities. The corresponding GNLSE is not integrable, unlike the classical NLS equation [19,20], but it admits a Lagrangian structure. The latter property suggests constructing solutions by means of the Anderson´s variational approximation (VA) [21,22], which is used, in various forms, in studies of optical solitons [23-31], see also reviews in Refs. [32] and [33]. The existence of the Lagrangian makes it also possible to apply the Noether´s theorem to identify dynamical invariants of GNLSE [33].

The rest of the paper is structured as follows. In Sec. 2 we derive the GNLSE modelling the propagation of light in the setting under the consideration. In Sec. 3 we introduce the Lagrangian for the GNLSE and elaborate the VA, using two different *ansätze* (trial forms of the approximate solution). We thus derive two sets of Euler-Lagrange equations corresponding to these *ansätze*. They involve effective Lagrangians which include the dilogarithm function $Li_2(z)$, where $z$ is the propagation distance (evolution variable). With the help of Noether´s theorem, we obtain dynamical invariants (conserved quantities) of the GNLSE. Moreover, the Euler-Lagrange equations obtained via the VA permit us to apply the Vakhitov-Kolokolov criterion in order to show that the soliton solutions of this equation are stable. In Sec. 4 we demonstrate that the VA produces both, stationary solitons and breathers. The existence of the breathers is confirmed by direct numerical simulations of the GNLSE. The simulations also reveal robust breathers with intrinsic vibrations featuring beats, as well as solutions exhibiting spontaneous splitting of the initial pulse in two or several secondary ones. The paper is concluded by Sec. 5.

## 2. The fundamental equation (GNLSE)

We consider a non-centrosymmetric photorefractive crystal, whose principal axes coincide with the axes $(x,y,z)$ of a rectangular coordinate system, the $x$ axis designating the crystal´s optical axis (the c-axis). A bias voltage is applied between the crystal´s boundaries in the $x$ direction. A polarized laser beam, with the electric field directed along $x$, propagates in the $z$ direction. Once a stationary state is reached, the beam´s electric field can be written as:

$$\vec{E}(x,z) = \varphi(x,z)\, e^{ikz}\hat{x},\tag{1}$$

with $k = k_0 n_e = 2\pi n_e/\lambda_0$, where $n_e$ is the unperturbed refractive index along the extraordinary c-axis, and $\lambda_0$ is the carrier wavelength in vacuum. Field $\vec{E}(x,z)$ satisfies the Helmholtz equation,

$$\nabla^2 \vec{E} + (k_0 n_e')^2 \vec{E} = 0 \,,\tag{2}$$

where $n_e'$ is the full refractive index along the c-axis, $(n_e')^2 = n_e^2 - n_e^4 r_{ef} E_{SC}$ [34, 35], where $r_{ef}$ is the electro-optic coefficient, and $\vec{E_{SC}} = E_{SC}\hat{x}$ is the space-charge field induced in the crystal. Substituting the form of $\vec{E}$, defined as per Eq. (1), in Eq. (2), and using the paraxial approximation, we obtain an equation of the NLS type:

$$i\frac{\partial \varphi}{\partial z} + \frac{1}{2k}\frac{\partial^2 \varphi}{\partial x^2} + \frac{k\Delta n}{n_e}\varphi = 0, \qquad (3)$$

$$\Delta n \equiv -n_e^3 r_{ef} E_{SC}/2. \qquad (4)$$

In a biased photovoltaic pyroelectric photorefractive crystal, the space-charge field has three contributions [6]:

$$E_{SC} = E_0 \frac{I_\infty + I_d}{I + I_d} + E_p \frac{I_\infty - I}{I + I_d} - E_{py}\frac{\lambda_c I}{I_d} \qquad (5)$$

where $I = n_e|\varphi|^2/(2\eta_0)$ is the beam´s intensity (with $\eta_0 = \sqrt{\mu_0/\varepsilon_0}$), $I_d$ is the dark irradiance, $I_\infty = I(x \to \pm\infty)$, and $E_0 = E(x \to \pm\infty)$. The background value $E_0$ is determined by the external voltage, $E_p$ depends on the polarization of light and photovoltaic coefficient, $E_{py}$ is the pyroelectric field, and coefficient $\lambda_c$ is determined by properties of the crystal. The pyroelectric contribution to $E_{SC}$ (the last term in Eq. (5)) is adopted according to the approximation introduced in [36].

We substitute (5) in the definition of $\Delta n$, given by Eq. (4) and the ensuing expression for $\Delta n$ is inserted in (3). Then, introducing scaled variables $\zeta = z/(kx_0^2)$, $s = x/x_0$, $U = n_e^{1/2}\varphi/(2P_0\eta_0)^{1/2}$, where $x_0$ and $P_0$ are characteristic units of the propagation length and intensity of the laser beam, Eq. (2) is transformed into:

$$i\frac{\partial U}{\partial \zeta} + \frac{1}{2}\frac{\partial^2 U}{\partial s^2} + \beta\gamma|U|^2 U - \theta\frac{\rho+1}{1+\gamma|U|^2}U - \alpha\frac{\rho-\gamma|U|^2}{1+\gamma|U|^2}U = 0, \qquad (6)$$

where $\rho = I_\infty/I_d$, $\alpha = E_p\tau$, $\beta = \lambda_c E_{py}\tau$, $\theta = E_0\tau$, $\tau = (k_0x_0)^2 n_e^4 r_{ef}/2$ and $\gamma = P_0/I_d$. Finally, we consider the illumination of the crystal corresponding to $I_\infty = \rho = 0$, hence Eq. (6) reduces to the GNLSE:

$$i\frac{\partial U}{\partial \zeta} + \frac{1}{2}\frac{\partial^2 U}{\partial s^2} + \beta\gamma|U|^2 U - \theta\frac{U}{1+\gamma|U|^2} + \alpha\frac{\gamma|U|^2 U}{1+\gamma|U|^2} = 0. \qquad (7)$$

This equation predicts the propagation of stationary spatial solitons in the photorefractive waveguide [6].

Stationary soliton solutions to Eq. (7) with a real propagation constant $\kappa$ are looked for as:

$$U = \exp(i\kappa\zeta)\,V(s;\kappa), \qquad (8)$$

where $V(s)$ is a real function. In particular, the main characteristic of the soliton is its integral optical power,

$$P(\kappa) = \int_{-\infty}^{+\infty} V^2(s;\kappa)ds, \qquad (9)$$

see also Eq. (12) below.

### 3. The Lagrangian structure

In terms of the appropriate Lagrangian density,

$$\mathcal{L} = \frac{i}{2}\left(U^*U_\zeta - UU_\zeta^*\right) - \frac{1}{2}U_sU_s^* + \frac{\beta\gamma}{2}U^2(U^*)^2 + \alpha UU^* - \frac{1}{\gamma}(\alpha+\theta)\,ln(1+\gamma UU^*)\,, \quad (10)$$

Eq. (7) is written as the variational Euler-Lagrange equation:

$$\frac{\partial\mathcal{L}}{\partial U} - \frac{\partial}{\partial\zeta}\frac{\partial\mathcal{L}}{\partial U_\zeta} - \frac{\partial}{\partial s}\frac{\partial\mathcal{L}}{\partial U_s} = 0. \quad (11)$$

According to the Noether's theorem, the invariance of the action integral, corresponding to the Lagrangian density (10), with respect to translations along directions $\zeta$ and $s$, and an arbitrary shift of the phase of the complex amplitude $U$, implies the conservation of three dynamical invariants, *viz*., the Hamiltonian, momentum, and optical power (norm) [cf. Eq. (9)]:

$$H = \int_{-\infty}^{+\infty} \mathcal{H}\,ds, \quad M = \int_{-\infty}^{+\infty} \mathcal{M}\,ds, \quad P = \int_{-\infty}^{+\infty} \mathcal{P}\,ds, \quad (12)$$

with densities:

$$\mathcal{H} = -\frac{1}{2}U_sU_s^* + \frac{\beta}{2}\gamma|U|^4 + \alpha\left[|U|^2 - \frac{1}{\gamma}\ln(1+\gamma|U|^2)\right] - \frac{\theta}{\gamma}\ln(1+\gamma|U|^2), \quad (13)$$

$$\mathcal{M} = \frac{i}{2}\left(U^*U_\zeta + UU_s^* - UU_\zeta^* - U^*U_s\right) - \frac{1}{2}U_sU_s^* + \frac{\beta}{2}\gamma|U|^4$$

$$+ \alpha\left[|U|^2 - \frac{1}{\gamma}\ln(1+\gamma|U|^2)\right] - \frac{\theta}{\gamma}\ln(1+\gamma|U|^2)\,, \quad (14)$$

$$\mathcal{P} = |U|^2\,. \quad (15)$$

The Lagrangian representation of Eq. (5) suggests the possibility of applying the Anderson´s VA [21, 22] to predict the propagation of a solitary-wave input in the photorefractive crystal.

To apply the VA, one needs to choose an adequate variational ansatz for a soliton (strictly speaking, it is a *solitary wave*, rather than a soliton in an integrable system). The traditional Gaussian, similar to the one used in Refs. [21, 28-31] for NLS equations with polynomial nonlinearities, would not allow the necessary analytical calculation of the resulting integral of the nonpolynomial Lagrangian density (10). Instead, two *ansätze* lead to VA in a tractable analytical form. The first one is based on the hyperbolic secant:

$$U(\zeta,s) = A(\zeta)\text{sech}\left[\frac{s}{a(\zeta)}\right]\exp[i\{b(\zeta) + c(\zeta)s^2\}]. \quad (16)$$

where $A$, $a$, $b$, and $c$ are, respectively, the soliton´s amplitude, width, phase, and *chirp*. Substituting this ansatz in the Lagrangian density (8), we calculate the corresponding *averaged Lagrangian* as:

$$L = \int_{-\infty}^{\infty} \mathcal{L} \, ds, \qquad (17)$$

The result being:

$$L = -2aA^2 \left( \frac{db}{d\zeta} - \alpha \right) - \frac{\pi^2 a^3 A^2}{6} \left( \frac{dc}{d\zeta} + 2c^2 \right) - \frac{A^2}{3a} + \frac{2\beta\gamma aA^4}{3} - \frac{a}{\gamma}(\alpha + \theta)f(y), \qquad (18)$$

where $y = \gamma A^2$, and we define:

$$f(y) = -\frac{\pi^2}{6} - \text{Li}_2\left( -1 - 2y - 2\sqrt{y(1+y)} \right) - \text{Li}_2\left( -1 - 2y + 2\sqrt{y(1+y)} \right), \qquad (19)$$

with the *dilogarithm* $\text{Li}_2$ (also called the *Spence´s function*),

$$\text{Li}_2(t) = -\int_0^t \frac{\ln(1-\tau)}{\tau} d\tau \equiv \sum_{k=1}^{\infty} \frac{t^k}{k^2} \qquad (20)$$

[37], which is a special case of the *polylogarithm* $\text{Li}_n$ for $n = 2$. Note that the arguments of $\text{Li}_2$ in Eq. (19) are negative, in agreement with the fact that the integral definition (20) is well defined for $t < 1$. Note that the dilogarithm appeared only in few previous works dealing with optical solitons [38, 39]. Therefore, it is worthy to note that this special function naturally arises in the study of solitons in photorefractive media, in the present context.

The averaged Lagrangian (18), gives rise to Euler-Lagrange equations for the variational parameters $A(\zeta)$, $a(\zeta)$, $b(\zeta)$ and $c(\zeta)$, which are defined in ansatz (16):

$$aA^2 = \text{const}, \qquad (21)$$

(this equation implies, as usual, the power conservation),

$$\frac{da}{d\zeta} = 2ac, \qquad (22)$$

$$\frac{db}{d\zeta} = \alpha - \frac{1}{3a^2} + \frac{5\beta\gamma A^2}{6} + \frac{(\alpha + \theta)}{4\gamma A^2}f(y) - \frac{3(\alpha + \theta)}{4}g(y), \qquad (23)$$

$$\frac{dc}{d\zeta} = -2c^2 + \frac{2}{\pi^2 a^4} - \frac{2\beta\gamma A^2}{\pi^2 a^2} - \frac{3(\alpha + \theta)}{\pi^2 a^2 \gamma A^2}f(y) + \frac{3(\alpha + \theta)}{\pi^2 a^2}g(y), \qquad (24)$$

where:

$$g(y) \equiv \frac{\mathrm{d}f(y)}{\mathrm{d}y}$$

$$= \frac{\left(-2 + \frac{1+2y}{\sqrt{y+y^2}}\right)\ln\left(2 + 2y - 2\sqrt{y+y^2}\right)}{-1 - 2y + 2\sqrt{y+y^2}}$$

$$+ \frac{\left(-2 - \frac{1+2y}{\sqrt{y+y^2}}\right)\ln\left(2 + 2y + 2\sqrt{y+y^2}\right)}{-1 - 2y - 2\sqrt{y+y^2}} \ . \tag{25}$$

The primary use of the Euler-Lagrange equations (21)-(24) is to predict the evolution of the solitary wave. These equations also make it possible to apply the Vakhitov-Kolokolov (VK) criterion [40-44] to predict the stability of the soliton solutions of the system (21)-(24).

For stationary solutions (alias *fixed points*) of the variational equations we set $\frac{da}{d\zeta} = 0$ in Eq. (22), which implies, as usual, that the chirp must vanish, $c = 0$. Then, setting $\frac{dc}{d\zeta} = 0$ in Eq. (24), leads to a certain VA-predicted relation between the amplitude and the width in the stationary solution:

$$a = H(A) \tag{26}$$

where the form of the function $H(A)$ can be obtained from Eq. (24).

Further, Eq. (23) produces the phase of the stationary solution as:

$$b(\zeta) = k\zeta, \tag{27}$$

where the propagation constant can be written as:

$$k = Q(A) \tag{28}$$

or, alternatively:

$$A = Q^{-1}(k) \tag{29}$$

taking into regard that Eq. (23) produces $k$ as a function of $A$ and $a$, and $a$ can be expressed in terms of $A$ as per Eq. (26).

To apply the VK criterion, one needs to obtain the sign of the derivative $dP/dk$, where $P$ is the optical power defined in Eq. (9). Using the form of the ansatz (16), and removing the width $a$ in favor of the amplitude $A$ pursuant to Eq. (26), we obtain:

$$P = \int_{-\infty}^{+\infty} |U|^2 \, ds = 2aA^2 = 2H(A)A^2 \tag{30}$$

Then, taking into account Eq. (29), we obtain:

$$\frac{dP}{dk} = \frac{dP}{dA}\frac{dA}{dk} = \frac{dP}{dA}\frac{1}{\frac{dk}{dA}} = \frac{dP}{dA}\frac{1}{Q'(A)} \tag{31}$$

where $Q'(A) = dQ/dA$.

As an example, Fig. 1 displays the VK derivative $dP/dk$, as obtained from Eq. (31) with the following (physically relevant) values of the coefficients in Eq. (7):

$$\alpha = -6.76, \quad \beta = 21.45, \quad \theta = -13.51, \quad \gamma = 1 \tag{32}$$

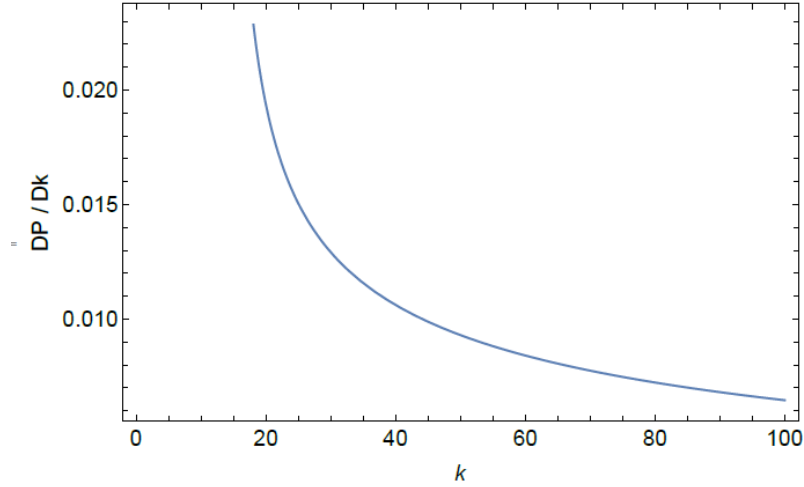

Fig. 1. Values of the VA-predicted VK derivative, $dP/dk$, for the family of stationary solitons based on ansatz (16), as produced by Eq. (31) with the coefficients fixed as per Eq. (32).

It is seen in Fig. 1 that the condition $\frac{dP}{dk} > 0$ holds for all values of $k$, and hence the VA-predicted family of stationary soliton solutions may be completely stable. In the following section we present some numerical solutions corroborating this prediction.

An alternative ansatz relies upon the square root of sech, instead of sech in Eq. (16), to approximate the soliton's shape:

$$U(\zeta, s) = A(\zeta) \sqrt{\operatorname{sech}\left[\frac{s}{a(\zeta)}\right]} \exp[i\{b(\zeta) + c(\zeta)s^2\}]. \tag{33}$$

where the parameters $A, a, b,$ and $c$ have the same meaning as in Eq. (16). Substituting this ansatz in the Lagrangian density (10) and calculating the integral (17), we obtain the averaged Lagrangian:

$$L = -a\pi A^2 \left(\frac{db}{d\zeta} - \alpha\right) - \frac{a^3\pi^3}{4} A^2 \left(\frac{dc}{d\zeta} + 2c^2\right) - \frac{\pi A^2}{16a} + \beta\gamma a A^4 - \frac{a}{\gamma}(\alpha + \theta)p(y), \tag{34}$$

where $y = \gamma A^2$ as above, and:

$$p(y) \equiv -\frac{\pi^2}{12} - 2\text{Li}_2\left(\frac{1}{-y + \sqrt{y^2-1}}\right) - 2\text{Li}_2\left(\frac{-1}{y + \sqrt{y^2-1}}\right). \tag{35}$$

Note that the averaged Lagrangian (34) also includes the dilogarithm $\text{Li}_2$. The occurrence of this special function in both averaged Lagrangians, (18) and (34), is due to the presence of the logarithmic term in the Lagrangian density (10). Consequently, it may be expected that $\text{Li}_2$ will appear whenever the VA is applied to other models which contain a logarithmic term in the Lagrangian.

Using the averaged Lagrangian (34) we derive the following system of Euler-Lagrange equations:

$$aA^2 = \text{const}, \tag{36}$$

$$\frac{da}{d\zeta} = 2ac, \tag{37}$$

$$\frac{db}{d\zeta} = \alpha - \frac{1}{8a^2} + \frac{5\beta\gamma A^2}{2\pi} + \frac{(\alpha+\theta)}{2\pi\gamma A^2}p(y) - \frac{3(\alpha+\theta)}{2\pi}q(y), \tag{38}$$

$$\frac{dc}{d\zeta} = -2c^2 + \frac{1}{4\pi^2a^4} - \frac{2\beta\gamma A^2}{\pi^3 a^2} - \frac{2(\alpha+\theta)}{\pi^3 a^2 \gamma A^2}p(y) + \frac{2(\alpha+\theta)}{\pi^3 a^2}q(y), \tag{39}$$

where:

$$q(y) \equiv \frac{dp(y)}{dy} = 2\frac{\left[\ln\left(\frac{1}{y-\sqrt{y^2-1}}+1\right) - \ln\left(\frac{1}{y+\sqrt{y^2-1}}+1\right)\right]}{\sqrt{y^2-1}}. \tag{40}$$

While Eqs. (36) and (37) are identical to (21) and (22), Eqs. (38) and (39) are different from their counterparts (23) and (24) produced by ansatz (16). Solutions of the system (36)-(39) are presented in the next section.

Having the Euler-Lagrange equations (36)-(39) we can apply the VK criterion to predict the stability of the soliton solutions. Following a procedure similar to that which was employed above to produce Fig. 1, and fixing the same parameters as in Eq. (32), we obtain a plot for the function $dP/dk$ which has a form which is almost identical to the one shown in Fig. 1. Therefore, ansatz (33) also predicts VK-stable solitons as stationary solutions of Eq. (7).

To close this section we would like to emphasize that the VA is a powerful tool in the study of solitons and breathers because it allows one to transform a complex nonlinear partial differential equation (NLPDE) into a system of first-order ordinary differential equations (ODEs), which is definitely much easier to solve than the underlying NLPDE. The Euler-Lagrange equations provided by the VA may also permit us to obtain, in a compact way, approximate results that would be quite difficult to obtain if we dealt directly with the original NLPDE. For example, in this section we were able to estimate the stability of the solitons of Eq. (7) by means of the VK criterion, because the variational equations (22)-(26) permitted us to calculate the sign of the derivative dP/dk.

## 4. Variational and full numerical solutions

## 4.1. Variational results produced by the sech ansatz (16)

We start the analysis by considering the same coefficients $\alpha$, $\beta$, $\theta$ and $\gamma$ from Eq. (32), that we used above to generate Fig. 1. Actually, these coefficients adequately represent the generic case.

First, Fig. 2 demonstrates the soliton's shape, as produced by ansatz (16) with parameters obtained from the numerical solution of the Euler-Lagrange equations (21)-(24), with the initial conditions corresponding to a fixed point of the equations, which is $A = \sqrt{10}$ and $a = 7/100$, the respective ansatz being

$$U(0, s) = \sqrt{10} \ \mathrm{sech} \ (100 \ s/7) \ . \tag{41}$$

The stationary propagation observed in Fig. 2 corroborates that this fixed point is a stable constant solution of equations (21)-(24).

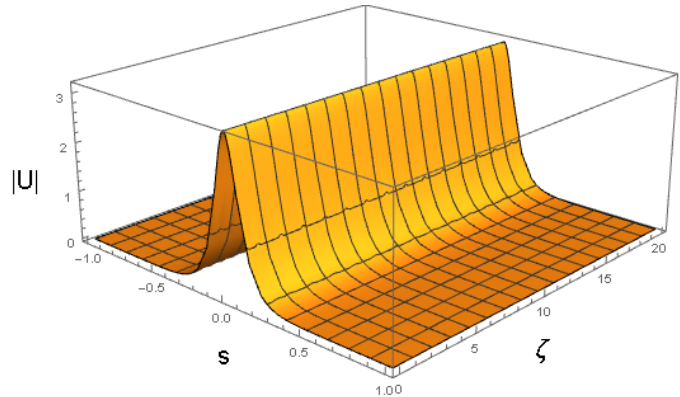

Fig. 2. $|U(s, \zeta)|$ as produced by ansatz (16), with the variational parameters obtained from the numerical solution of equations (21)-(24) with coefficients (32), and initial condition corresponding to input (41), which represents the fixed point of the equations. The values of coefficients (32) correspond to $E_{py} = 12.7 \ kV/cm$, $E_0 = -3.99518 \ kV/cm$, $E_p = -2 \ kV/cm$ and the physical parameters shown in Eqs. (47). The counterpart of this picture produced by full simulations of GNLSE (7) is presented below in Fig. 16.

Next, we take coefficients:

$$\alpha = 31.5965, \beta = 3.9049, \theta = -13.5328, \gamma = 1 \tag{42}$$

in Eq. (7), and initial parameters in ansatz (16) corresponding to the input:

$$U(0, s) = 4 \operatorname{sech}(8s), \tag{43}$$

*cf.* Eq. (41). The amplitude and width of the expression (43) do not correspond to a fixed point of Eqs. (22) and (24), hence the solution of the Euler-Lagrange equations (21)-(24) produces the solution shown in Figs. 3 and 4. It is a *breather*, *i.e.*, a solitary wave whose amplitude and width exhibit periodic oscillations along the $\zeta$-axis with period $\lambda \approx 0.123$.

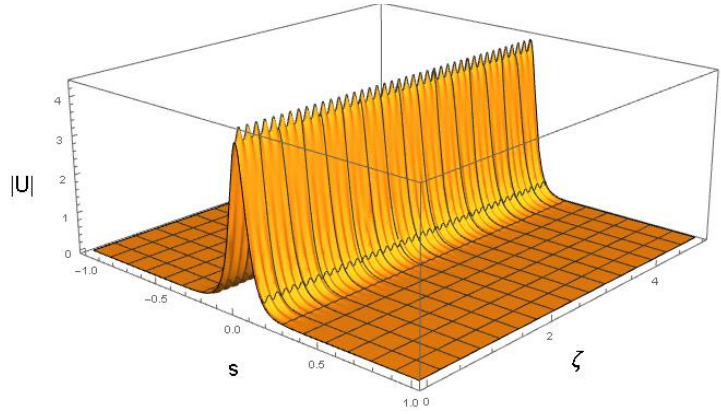

Fig. 3. The VA-predicted breather for the coefficients chosen as per Eq. (42), and the input corresponding to Eq. (43). The values of coefficients (42) correspond to $E_{py} = 2.31\ kV/cm$, $E_0 = -4.00035\ kV/cm$, $E_p = 9.34\ kV/cm$ and the physical parameters shown in Eqs. (47). The counterpart of this dynamical picture produced by full simulations of GNLSE (7) is presented below in Fig. 18.

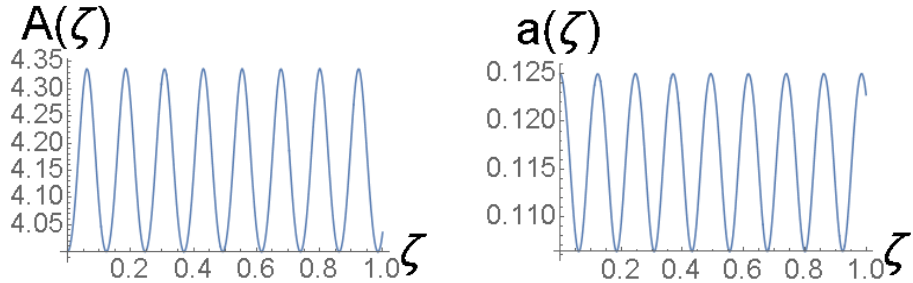

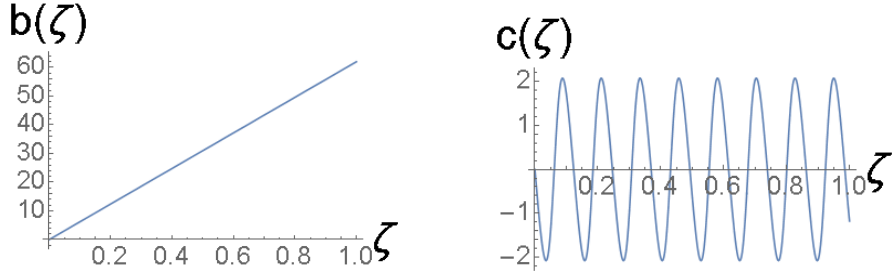

Fig. 4. Solutions of the Euler-Lagrange equations (21)-(24) for variational parameters $\{A(\zeta), a(\zeta), b(\zeta), c(\zeta)\}$, for the same case as shown in Fig. 3.

Another relevant example of breathers is produced by Eqs. (21)-(24) with a parameter set which differs from (42) by a smaller value of coefficient $\beta$ in front of the cubic term:

$$\alpha = 31.5965, \beta = 1.0149, \theta = -13.5328, \gamma = 1, \tag{44}$$

and initial values of the variational parameters corresponding to the following input:

$$U(0, s) = 4 \operatorname{sech}(6s), \tag{45}$$

*cf.* Eq. (43). Similar to the example displayed in Figs. 3 and 4, the amplitude and width of expression (45) do not correspond to a fixed point of Eqs. (22) and (24). In this case, the solution of the Euler-Lagrange equations (21)-(24) produces the results presented in Figs. 5 and 6. They exhibit oscillations of the breather with period $\lambda \approx 0.411$, which is larger by a factor $\simeq 3.34$ than in Figs. 3 and 4.

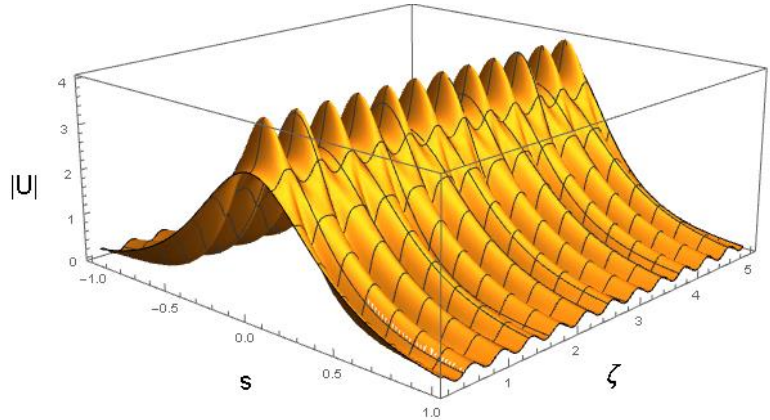

Fig. 5. The same as in Fig. 3, but for parameters (44) and the input corresponding to Eq. (45). The values of coefficients (44) correspond to $E_{py} = 0.6 \, kV/cm$, $E_0 = -3.99518 \, kV/cm$, $E_p = 9.34 \, kV/cm$ and other physical parameters shown in Eqs. (47). The counterpart of this dynamical picture produced by full simulations of GNLSE (7) is presented below in Fig. 19.

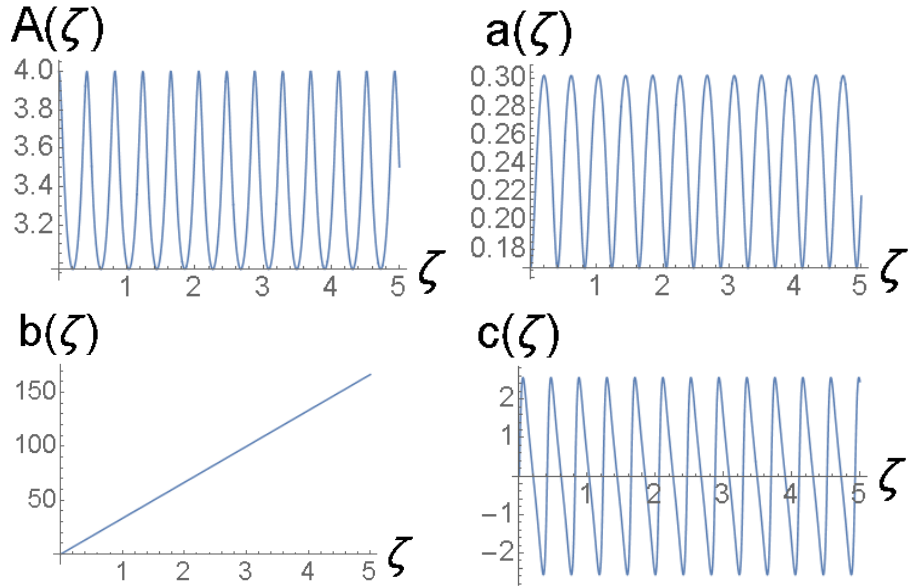

Fig. 6. Solutions of the Euler-Lagrange equations (21)-(24) for variational parameters $\{A(\zeta), a(\zeta), b(\zeta), c(\zeta)\}$ for the same case as shown in Fig. 5.

Lastly, Figs. 7 and 8 demonstrate an example of a robust breather featuring much slower oscillations, with period $\lambda \approx 1.888$, which is larger by a factor $\simeq 4.6$ than in Figs. 5 and 6. This solution of Eqs. (21)-(24) was obtained for a parameter set:

$$\alpha = -6.7658, \beta = 14.078, \theta = -13.5153, \gamma = 1 \qquad (46)$$

and the same initial input (41) as used above, which in this case does not represent a fixed point of Eqs. (22) and (24) either.

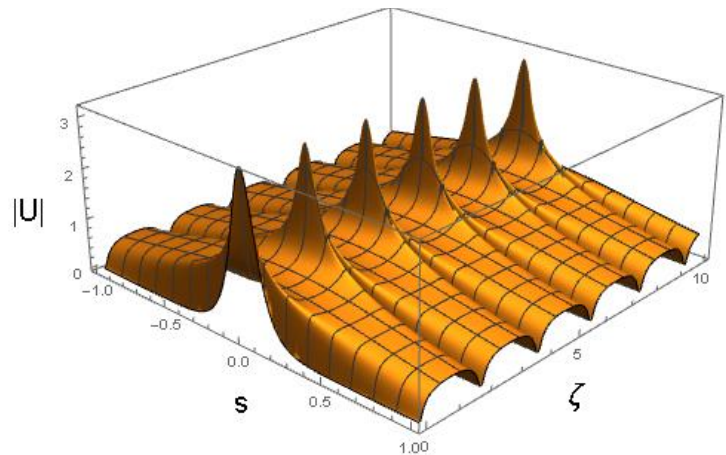

Fig. 7. The same as in Fig. 3, but for parameters (46) and the input corresponding to Eq. (41). The values of coefficients (46) correspond to $E_{py} = 8.32\ kV/cm$, $E_0 = -3.99518\ kV/cm$, $E_p = -2\ kV/cm$ and other physical parameters shown in Eqs. (47).

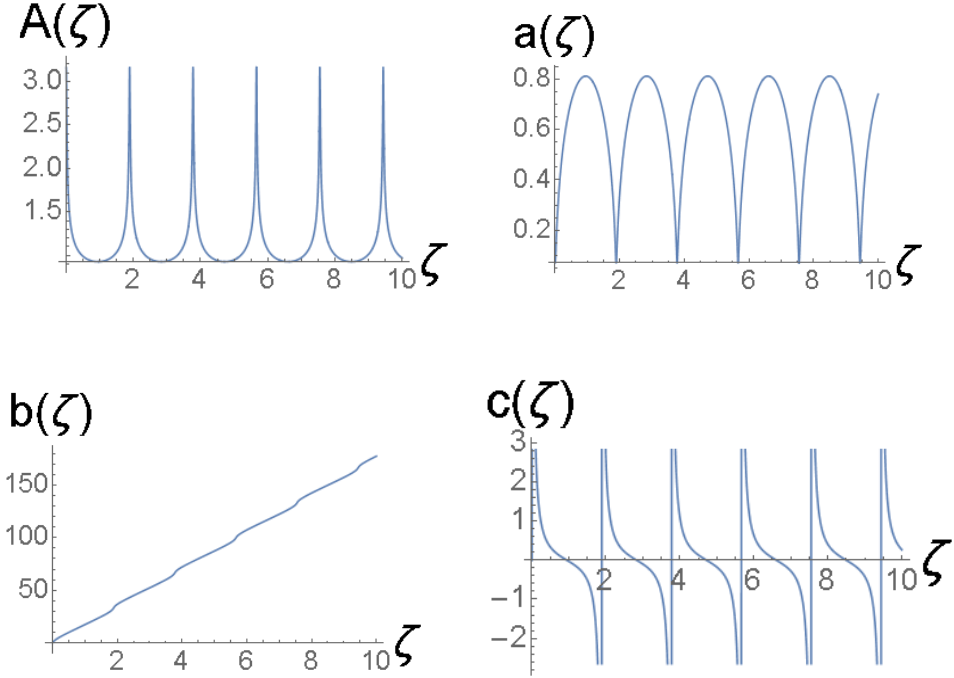

Fig. 8. Solutions of the Euler-Lagrange equations (21)-(24) for variational parameters $\{A(\zeta), a(\zeta), b(\zeta), c(\zeta)\}$ for the same case as shown in Fig. 7.

Thus, depending on values of coefficients $\alpha$, $\beta$ and $\theta$, in Eq. (7) and the form of the initial condition, VA predicts breathers with very different oscillation periods. The values of the coefficients $\alpha$, $\beta$ and $\theta$ used to obtain the solutions shown in Figs. 2-8 are translated, using the material characteristics given [6], into the following values of the physical parameters of the photorefractive crystal:

$$
\begin{array}{ll}
x_0 = 20\mu m & k_0 = \frac{2\pi}{\lambda_0} = \frac{2\pi}{405} nm^{-1} \\
n_e = 2.2 & r_{ef} = 30 \times 10^{-12} m/V \\
\lambda_c = 0.5 & \tau = 3.3829 \times 10^{-5} m/V
\end{array}
\tag{47}
$$

### 4.2. Variational results produced by the $\sqrt{\mathbf{sech}}$ ansatz (33)

Next, we produce VA predictions based on the alternative analytically tractable ansatz (33), with the evolution of the variational parameters governed by the Euler-Lagrange equations (36)-(39). To begin with, in Fig. 9 we produce a stationary soliton generated by initial values of the variational parameters corresponding to the input:

$$U(0, s) = \sqrt{10 \operatorname{sech}(13.11\, s)} \tag{48}$$

whose amplitude and width correspond to a fixed point of Eqs. (37) and (39), and the respective set of coefficients in Eq. (7) being:

$$\alpha = -6.7658,\ \beta = 7.7976,\ \gamma = 1,\ \theta = -13.5153 \tag{49}$$

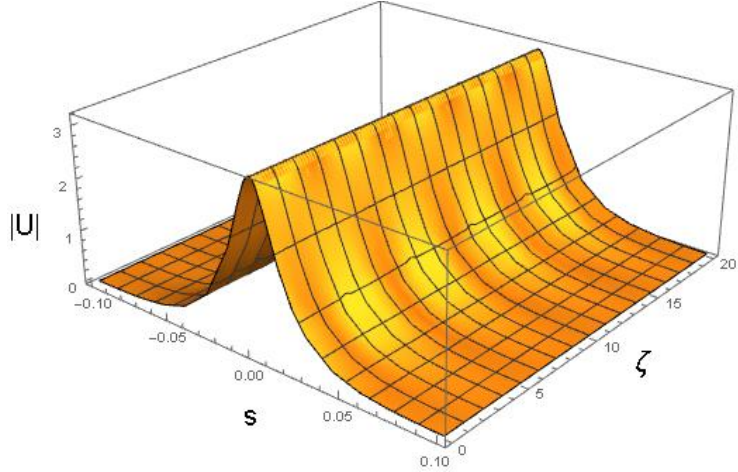

Fig. 9. Soliton obtained with ansatz (33). In this case the variational parameters evolve according to the Euler-Lagrange equations (36)-(39), with the coefficients shown in Eq. (49), and the initial condition given in Eq. (48). The values of coefficients (49) correspond to $E_{py} = 4.61\ kV/cm$, $E_p = -2\ kV/cm$, $E_0 = -3.99518\ kV/cm$, and the physical parameters shown in Eqs. (47).

Next, to produce a breather shown in Figs. 10 and 11, we take the same input (48), but with different coefficients in Eqs. (7) and (36)-(39):

$$\alpha = -6.7658,\ \beta = 7.2976,\ \theta = -13.5153,\ \gamma = 1\ . \tag{50}$$

In this case, parameters of input (48) do not correspond to a fixed point of Eqs. (37) and (39), and consequently the simulations of the variational equations (36)-(39) produce a breather with a period of stable oscillations $\lambda \approx 0.24$.

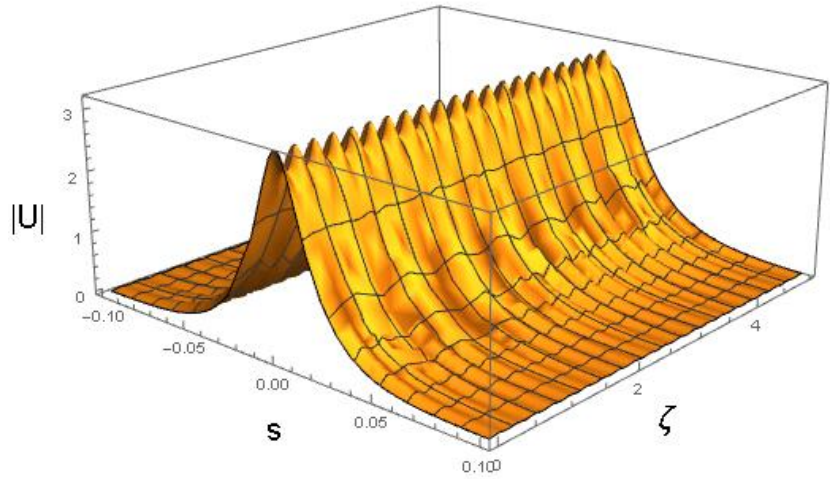

Fig. 10. VA-predicted breather obtained with the ansatz (33). It is plotted with the variational parameters evolving according to the Euler-Lagrange equations (36)-(39), with the coefficients chosen as per Eq. (50) and the initial condition corresponding to Eq. (48). The values of coefficients (50) correspond to $E_{py} = 4.31 \, kV/cm$, $E_p = -2 \, kV/cm$, $E_0 = -3.99518 \, kV/cm$, and other physical parameters shown in Eqs. (47).

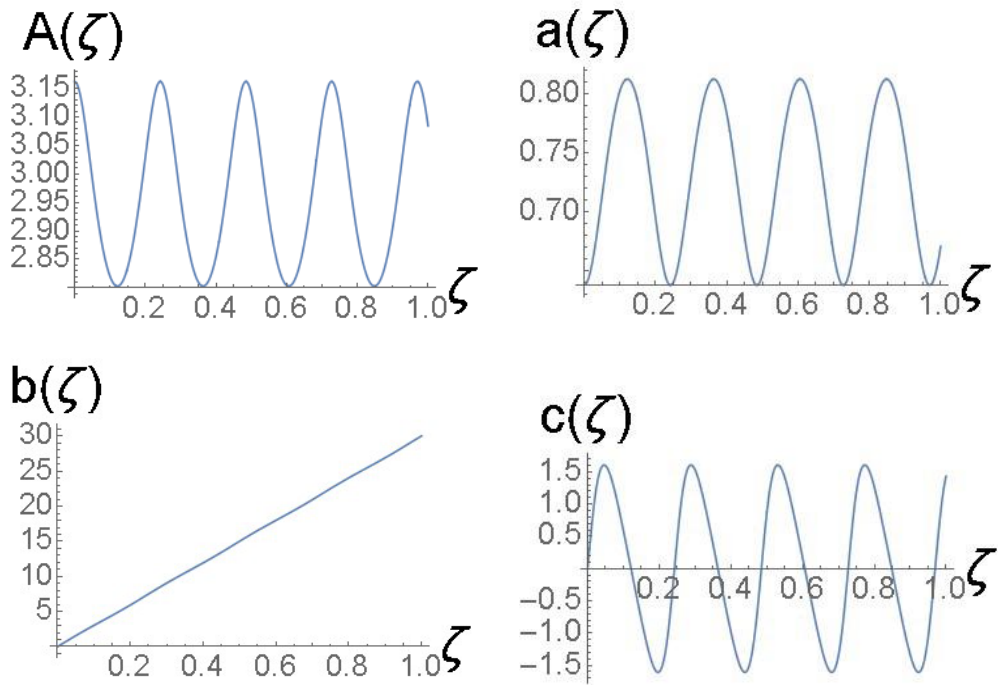

Fig. 11. Solutions of the Euler-Lagrange equations (36)-(39) for variational parameters $\{A(\zeta), a(\zeta), b(\zeta), c(\zeta)\}$, for the same case as shown in Fig. 10.

Further, the same input (47) and the same set of coefficients (49), but with a smaller value of the coefficient in front of the cubic term, $\beta = 6.6$ (in this case, the input does not represent a fixed point of Eqs. (37) and (39) either), produce the breather shown in Figs. 12 and 13. Its oscillation period, $\lambda \approx 0.56$, is larger by a factor $\approx 2.3$ than in Figs. 10 and 11.

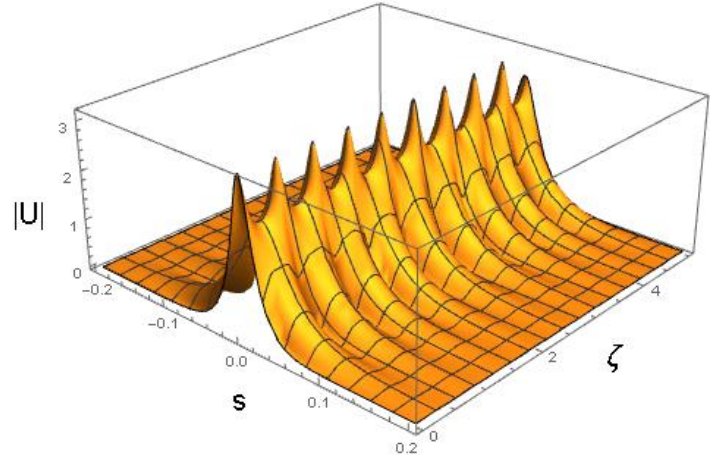

Fig. 12. The same as in Fig. 10, except for $\beta = 6.6$. This value, and coefficients $\alpha$, $\theta$ and $\gamma$ given in Eq. (50), correspond to $E_{py} = 3.9\ kV/cm$, $E_p = -2\ kV/cm$, $E_0 = -3.99518\ kV/cm$, and other physical parameters shown in Eqs. (47).

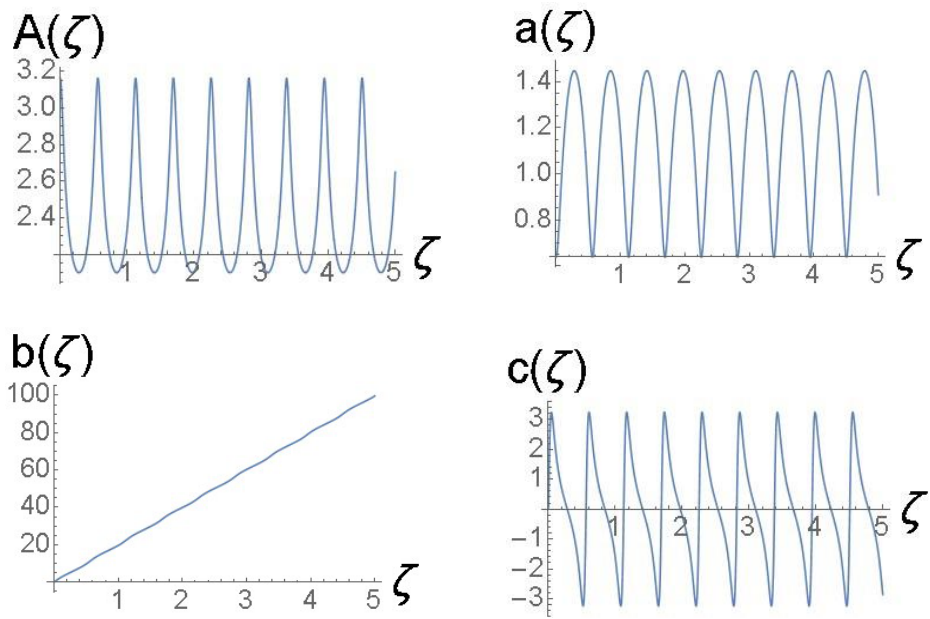

Fig. 13. Solutions of the Euler-Lagrange equations (36)-(39) for variational parameters $\{A(\zeta), a(\zeta), b(\zeta), c(\zeta)\}$, for the same case as shown in Fig. 12.

Lastly, a stable breather with a very large oscillation period, $\lambda \approx 5.8$, which is $\approx 24$ times larger than in Figs. 12 and 13, is predicted by VA based on the input:

$$U(0, s) = \sqrt{10 \text{sech}(22.38\, s)} \qquad (51)$$

for coefficients:

$$\alpha = -6.7658, \beta = 1.1976, \theta = -13.5153\,, \gamma = 1 \qquad (52)$$

in Eqs. (36)-(39). This slowly oscillating breather (once again, initiated by an input which does not correspond to the fixed point of Eqs. (37) and (39)), is displayed in Figs. 14 and 15.

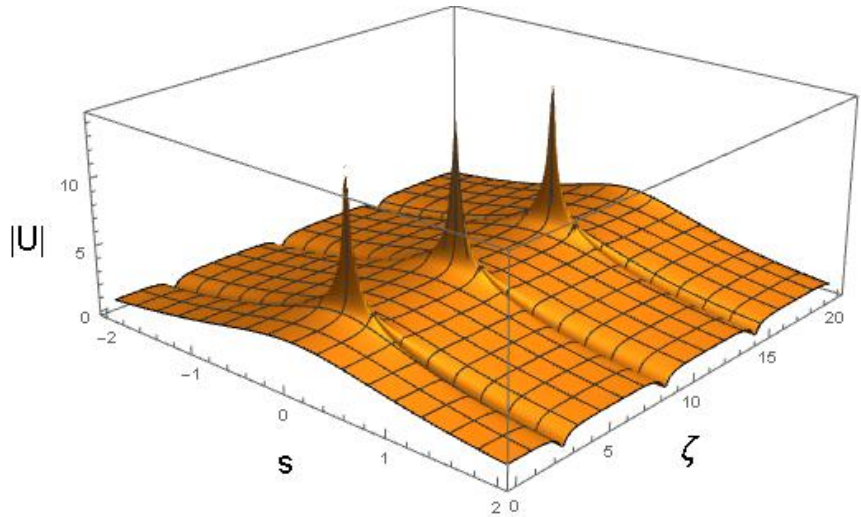

Fig. 14. Breather produced by ansatz (33). In this case, the variational parameters evolve according to the Euler-Lagrange equations (36)-(39), with the coefficients shown in Eq. (52) and the initial condition given in Eq. (51). The values of coefficients (52) correspond to $E_{py} = 0.71\, kV/cm$, $E_p = -2\, kV/cm$, $E_0 = -3.99518\, kV/cm$, and other physical parameters shown in Eqs. (47).

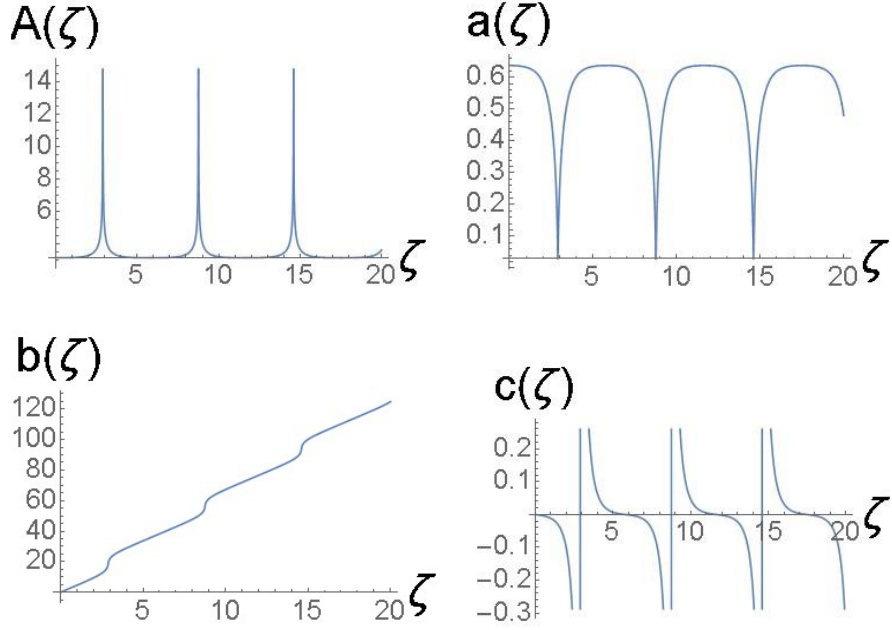

Fig. 15. Solutions of the Euler-Lagrange equations (36)-(39) for variational parameters $\{A(\zeta), a(\zeta), b(\zeta), c(\zeta)\}$, for the same case as shown in Fig. 14.

The values of coefficients $\alpha$, $\beta$ and $\theta$ used to obtain the solutions shown in Figs. 9-15 are translated, using the material characteristics given in book [18], into the values of the physical parameters of the photorefractive crystal given in Eqs. (47), along with $E_p = -2\ kV/cm$ and $E_0 = -3.99518\ kV/cm$.

### 4.3. Direct simulations of the evolution of solitons and breathers

The predictions of VA reported above have been verified by means of the full numerical solution of Eq. (7). These numerical solutions were obtained by using a 4th order Runge-Kutta algorithm to advance along $\zeta$, and approximating the second derivative $U_{ss}$ with finite differences including first and second neighbours. To begin with, we solved Eq. (7) using the same coefficients [Eq. (32)] and initial conditions [Eq. (41)] that we used to produce the variational solution shown in Fig. 2. The corresponding numerical solution of Eq. (7) is displayed in Fig. 16. It is seen that the full simulations produce a stable soliton in perfect agreement with the VA prediction.

The surfaces seen in Figs. 2 and 16 look quite similar, but in order to appreciate to what extent these surfaces are really similar, in Fig. 17 we show the form of the profile $|U(s, \zeta = 20)|$ given by the numerical solution of Eq. (7) [black squares], and the profile obtained from the variational solution [the thin continuous line]. It is seen that the agreement between the two solutions is excellent.

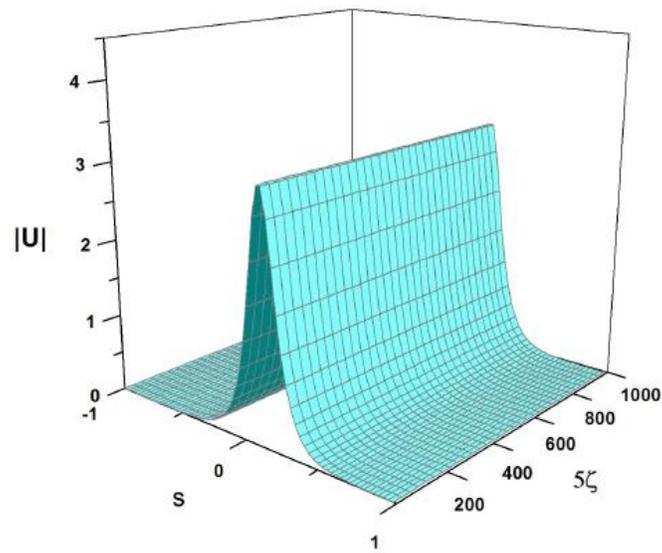

Fig. 16. Results of the numerical simulations of Eq. (7) with the same coefficients [see Eq. (32)] and the same input [Eq. (41)] which were used above to produce the VA solution displayed in Fig. 2.

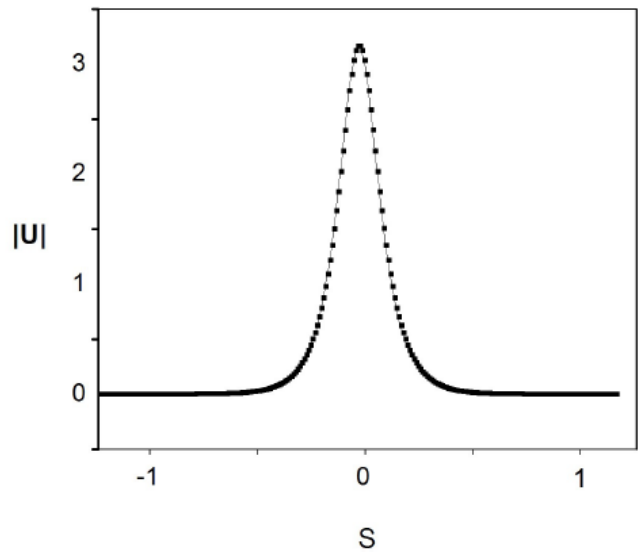

Fig. 17. Profiles $|U(s, \zeta = 20)|$ obtained from the numerical solution of Eq. (7) shown in Fig. 16 [black squares], and the variational solution shown in Fig. 2 [the thin line].

Next, in Figs. 18 and 19 we show numerical solutions of Eq. (7) which corroborate the existence of robust breathers which are quite close to those predicted above in Figs. 3 and 5 for the same parameters and initial conditions. It is worthy to note that the VA-predicted and numerically found oscillation periods are very close.

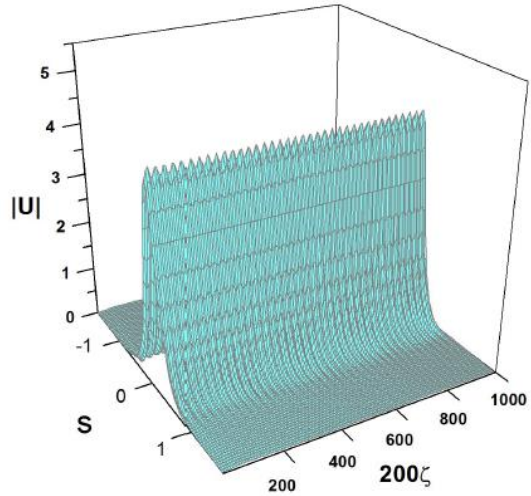

Fig. 18. Results of the numerical simulations of Eq. (7) with the same coefficients [see Eq. (42)] and the same input [Eq. (43)] which were used above to produce the VA solution displayed in Fig. 3.

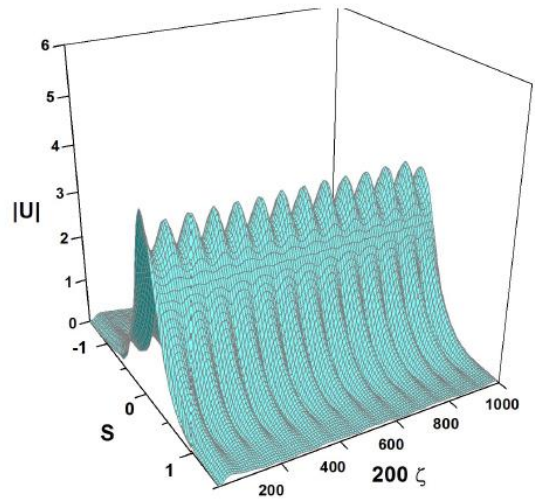

Fig. 19. Results of the numerical simulations of Eq. (7) with the same coefficients [see Eq. (44)] and the same input [Eq. (45)] which were used above to produce the VA solution displayed in Fig. 5.

To appreciate to what extent the variational solution shown in Fig. 5 is similar to the direct numerical solution shown in Fig. 19, in Fig. 20 we present the profiles $|U(s, \zeta = 1)|$ obtained from the variational solution shown Fig. 5 [with a thin line], and the profile obtained from the direct numerical solution of Eq. (7) [with black dots]. It is seen that the agreement between both profiles is very good, thus confirming that the variational method indeed produces solutions which are close to the direct numerical solutions of Eq. (7).

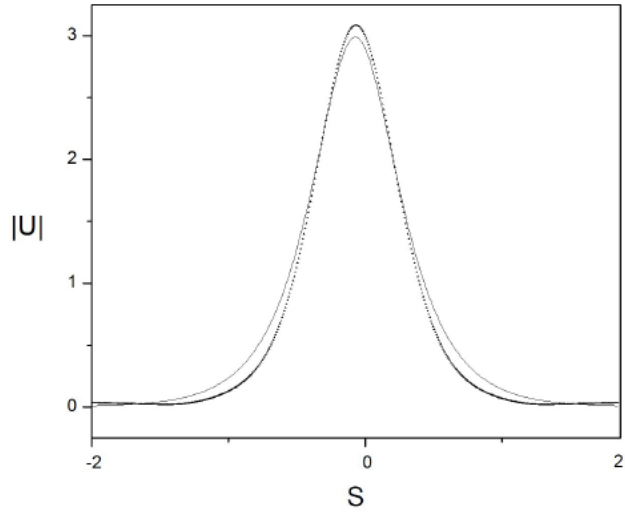

Fig. 20. Profiles $|U(s, \zeta = 1)|$ obtained from the numerical solution of Eq. (7) shown in Fig. 19 [black dots], and the variational solution shown in Fig. 5 [the thin line].

The breathing solutions shown in Figs. 18 and 19 were produced by using initial conditions which were very close to exact breathers. A natural question is what will happen if one uses an input which is not too close to an exact breather. To address this issue, in Fig. 21 we show the numerical solution of Eq. (7) obtained with the input:

$$U(s, 0) = 2.958 \text{ sech } (100s/7) \tag{53}$$

and the coefficients given in Eq. (46). It is observed that the input evolves into a breather, even though the initial amplitude of the pulse is significantly higher than the breather´s amplitude. This result implies that even if the breather is significantly perturbed (by increasing its amplitude and reducing its width), the pulse returns to the breather´s shape, indicating that this breather is a robust solution.

All the breather solutions produced by the simulations of Eq. (7) demonstrates their robustness against decay into small-amplitude radiation waves. Strictly speaking, the emission of radiation, in the form of slow decay of an excited state of a fundamental soliton [45] is generated by any equation of the NLS type (a well-known fact is that the integrable NLS equation with the cubic self-focusing gives rise to exact breather solutions in the form of $N$-solitons, which are produced by the input in the form of a fundamental soliton multiplied by an arbitrary integer $N = 2, 3, 4, \ldots$ [46]). Nevertheless, in our simulations the decay is a very weak effect, which remains virtually invisible.

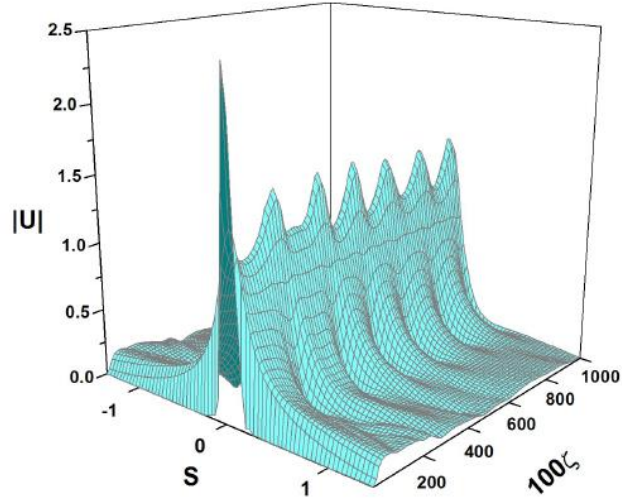

Fig. 21. Numerical solution of Eq. (7) for the coefficients shown in (46), and the initial condition (53).

We can use the numerical solutions of Eq. (7) to corroborate that the VK criterion predicts the stability of the soliton solutions of Eq. (7). To this end, we looked for solitons corresponding to the same set of coefficients $\{\alpha, \beta, \gamma, \theta\}$ as chosen in Eq. (49), and four different values of the propagation constant: $k = 21.99, 31.42, 62.83$ and $100.53$. In each case, we computed the integral power $P$ [see Eq. (12)], as shown in Fig. 22. This figure shows that dependence $P(k)$ has a *positive* slope (*i.e., $dP/dk > 0$*), hence the VK criterion indicates that these solitons are stable solutions, which is corroborated by the simulations of Eq. (7).

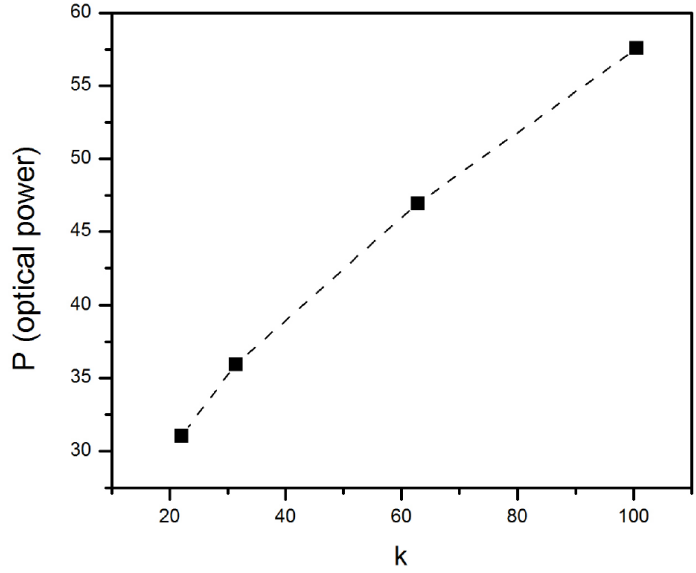

Fig. 22. Optical power $P$ vs. the propagation constant $k$ of the soliton solutions of Eq. (7) with the coefficients given in Eq. (49).

Figures (17)-(19) show that the separation $\lambda$ between the breathers´ peaks changes if we modify the coefficients of Eq. (7), and it is worthy to note that that the VA presented in Figs. 3, 5 and 7 predicts similar changes in $\lambda$. To evaluate to what extent the VA-predicted changes in $\lambda$ are similar to those observed in the direct numerical solutions of Eq. (7), we have calculated how $\lambda$ depends on $\beta$ according to the direct numerical solutions of Eq. (7) and, on the other hand, according to the variational equations (21)-(24). We focus here on the parameter $\beta$, as this is the coefficient associated to the cubic nonlinearity of Eq. (7), which is responsible for the creation and propagation of solitons in NLS-like equations.

In Fig. 23 we display the form of function $\lambda(\beta)$ according to the direct numerical solutions of Eq. (7) [black squares], and according to the variational equations [stars]. All the results shown in this figure were obtained using the coefficients $\alpha = 31.60$, $\theta = -13.53$, $\gamma = 1$ in Eq. (7), and the initial condition defined in Eq. (45).

Figure 23 shows that both procedures [performed using direct numerical solutions of Eq. (7), or the variational equations], show that $\lambda$ decreases as $\beta$ increases. This behaviour can be understood if we observe that the definition of $\beta$ [below Eq. (6)], combined with Eqs. (4) and (5), imply that $\Delta n$ will *increase* if $\beta$ increases. And if $\Delta n$ (the refractive index change due to the photorefractive effect) increases, the light velocity in the crystal decreases. Consequently, if $\Delta t$ is the time interval between two successive maxima in the light intensity of a breather, the distance $v\Delta t$, advanced by the light during this time interval, diminishes if the velocity of light ($v$) decreases. In other words, $\lambda = v\Delta t$ *will decrease*. This is the physical interpretation of Fig. 23. It should be mentioned, however, that a change in $\beta$ is not the only possible cause for a change in $\lambda$. In particular, it is known that light absorption may also change the value of $\lambda$ [47].

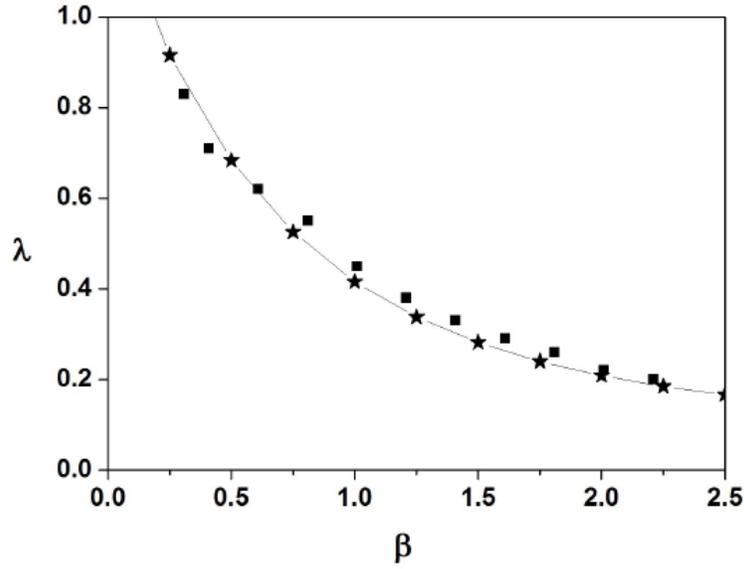

Fig. 23. Separation $\lambda$ between the breather´s peaks as a function of $\beta$. Black squares represent direct numerical solutions of Eq. (7), and the stars were obtained from the variational solution, using the ansatz (16). The input is taken as per Eq. (45), and other coefficients in Eq. (7) are: $\alpha = 31.60$, $\theta = -13.53$ and $\gamma = 1$.

The numerical solutions of Eq. (7) presented in this subsection, as well as the variational solutions presented in the subsections 4.1 and 4.2, show that Eq. (7) permits the propagation of solitons and breathers. Consequently, a natural question is if there is a criterion that permits one to forecast if a given solitary-wave input will propagate as a soliton, or it will evolve as a breather. In the case of the cubic NLS equation the integrability of this equation permits one to answer this question. More precisely: if we consider the input in the form $U(s, \zeta = 0) = A \operatorname{sech}(s/a)$, the Zakharov-Shabat equations predict the number of solitons that will be generated as the number of bound eigenstates in a linear problem (in a similar way as in the usual quantum mechanical problem with a given potential well). Namely, it is well-known [46] that, if the parameters of the input satisfy the inequality $Aa < 1/2$, the potential well has no bound states, and no solitons appear. In the interval $1/2 < Aa < 3/2$ the eigenvalue problem has exactly one bound state, hence exactly one soliton appears. However, when $Aa > 3/2$, the second bound state appears, which implies the creation of the superposition of two solitons, *i.e.*, precisely the simplest breather. In other words, in the integrable-NLS case, the character of the output (*i.e.*, the generation of a soliton or a breather), is determined by the size of the product $Aa$. For a nonintegrable equation such as Eq. (7), there are no exact results like that, although the situation remains qualitatively similar: for relatively small values of $Aa$, the input is able to maintain the self-trapping of a single soliton, but for larger values of $Aa$, the input cannot avoid the generation of at least one additional soliton, whose interaction with the original soliton builds a breather. It should be mentioned, however, than the determination of precise criteria which may be helpful to predict if a given solitary-wave initial condition for Eq. (7) will evolve as a soliton, or as a breather, is outside the scope of this article.

### 4.4. Breathers modulated by beats

The solutions produced by the simulations of Eq. (7) in Figs. 18, 19 and 21 may be considered as *regular* breathers, as they exhibit strictly periodic oscillations. On the other hand, non-integrable systems (such as the NLS equation combining the cubic self-defocusing and harmonic-oscillator trapping potential [48]) may give rise, under special conditions, to dynamical states featuring quasi-periodic oscillations. In accordance with this expectation, Eq. (7) gives rise, in addition to the regular breathers presented above, to more sophisticated spatially self-trapped modes in the form of breathers featuring long-period beats, such as one displayed in Fig. 24. It corresponds to coefficients:

$$\alpha = -6.7658, \quad \beta = 67.6584, \quad \theta = -13.2643, \quad \gamma = 1 \tag{54}$$

in Eq. (7), and the input:

$$U(s, \zeta = 0) = \sqrt{10} \text{ sech } (9.57s). \tag{55}$$

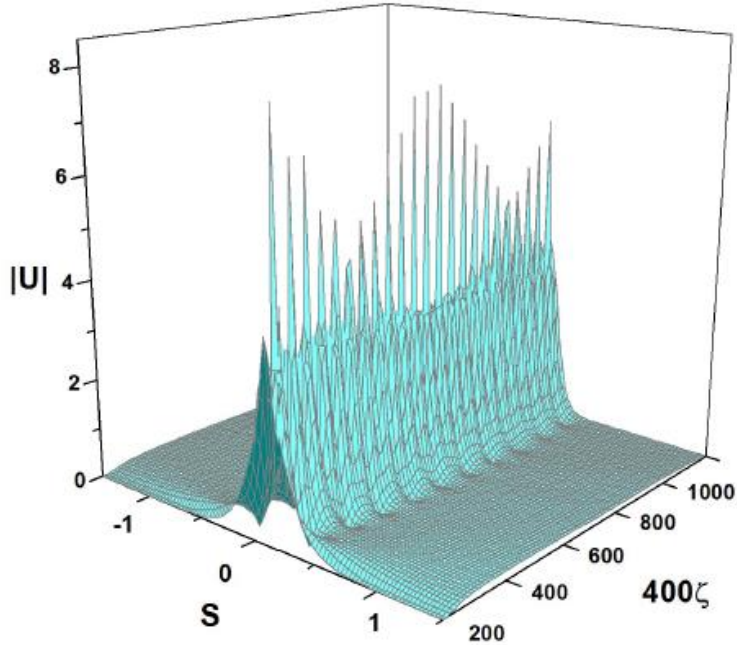

Fig. 24. Beating breather, obtained as a numerical solution of Eq. (7) with parameters (54) and input (55). The values of coefficients (54) correspond to the physical parameters given in Eqs. (47), $E_{py} = 40 \, kV/cm$, $E_0 = -3.92098 \, kV/cm$ and $E_p = -2 \, kV/cm$.

To identify the character of the beats exhibited by the breather displayed in Fig. 24, it is relevant to calculate the power spectrum of oscillations of their amplitude, $\left|\widetilde{U}(\nu)\right|^2$, where $\widetilde{U}(\nu)$ is the Fourier transform of $U(\zeta, s = 0)$, $\nu$ standing for the respective spatial frequency.

The power spectrum for the amplitude oscillations of the breather shown in Fig. 24 is plotted in Fig. 25. In this case, the Fourier transform of $|U(\zeta, s = 0)|$ was computed using $2^{14}$ values of this function of $\zeta$, corresponding to 16384 equally spaced points in the interval of $0 \leq \zeta \leq 7.5$. As $\zeta$ is a dimensionless variable, the Fourier transform of $|U(\zeta, s = 0)|$ is a function of a *dimensionless spatial frequency*. It is observed that the discrete spectral components at frequencies $\pm 1.09, \pm 2.19, \pm 3.28$ represent anharmonic oscillations of the breather, while the dc component concentrated around the zero frequency accounts for long-period beats.

The results presented in this section, and in the previous ones, show that breathers in photorefractive crystals constitute a subject deserving further investigation. In this connection, it is relevant to mention that the study of breathers in other systems, such as linear and nonlinear phononic crystals, and granular crystals [49-51], and the study of discrete breathers [52-54], are also interesting fields.

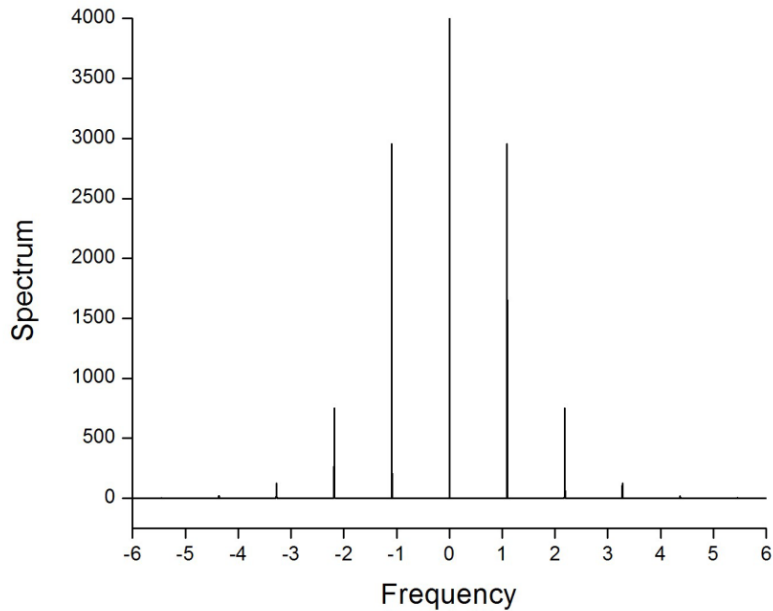

Fig. 25. The power spectrum of the amplitude function $|U(\zeta, s = 0)|$ corresponding to the breather shown in Fig. 24.

## 4.5 Splitting breathers

Another example of a breather with slowly modulated intrinsic oscillations is displayed in Fig. 26. It is obtained as a numerical solution of Eq. (7) with the same values of coefficients as in Eq. (53) and the input:

$$U(s, \zeta = 0) = \sqrt{10} \text{ sech } (8.70s). \qquad (56)$$

Unlike the dynamical scenarios demonstrated above, this breather splits in two parts: a small solitary wave that moves to the left (towards $s < 0$), and the principal one, which moves to the right with a small velocity. While the small solitary wave which moves to the left is not seen in Fig. 26 because it is located behind the principal wave, it is visible under the different angle used in Fig. 27. Actually, splitting is one of generic outcomes of the evolution of chirped pulses in equations of the NLS type [55, 56].

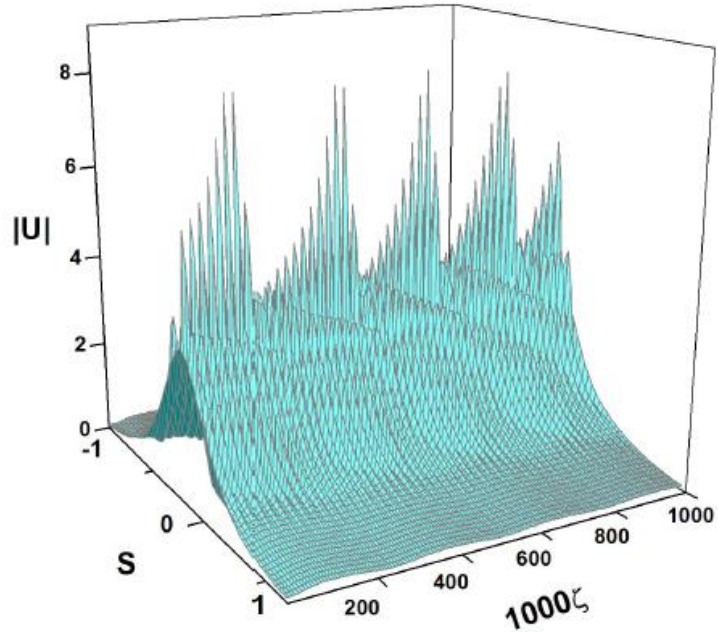

Fig. 26. Numerical solution of Eq. (7) with values of parameters (54) and input (56). These coefficients correspond to the physical parameters shown in Eqs. (47), and $E_{py} = 40\ kV/cm$, $E_0 = -3.92098\ kV/cm$ and $E_p = -2\ kV/cm$.

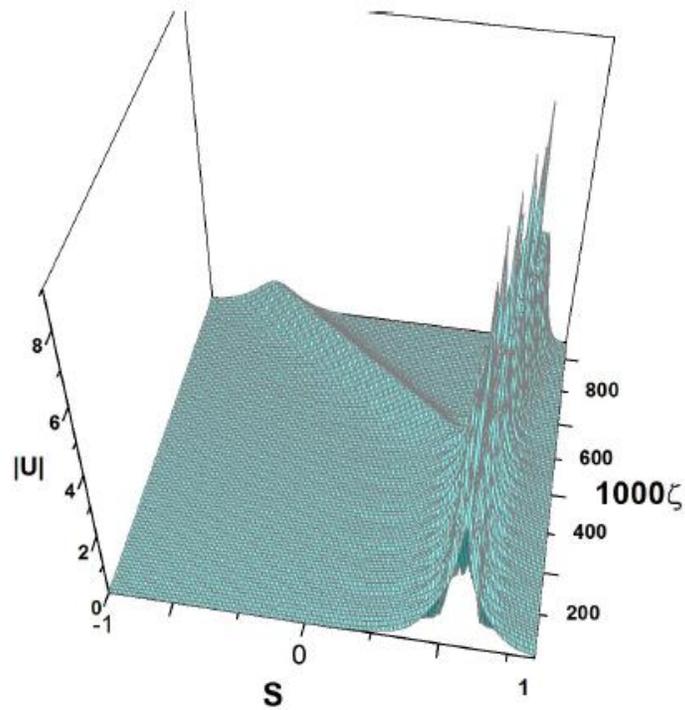

Fig. 27. The same as in Fig. 26, but observed from a different direction.

Splitting of a single pulse into multiple solitary waves is possible too. A typical example of this type is displayed in Fig. 28, where the initial pulse is fragmented into five solitons with weak internal oscillations. In this case, the coefficients of Eq. (7) are the same as in Fig. 24, *i.e.*, they are given as per Eq. (54), and the input is:

$$U(s, \zeta = 0) = \sqrt{10} \text{ sech } (4.50s). \tag{57}$$

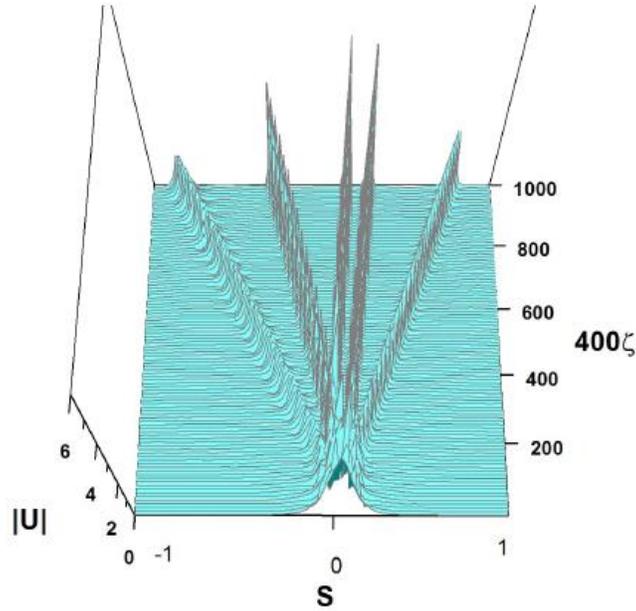

Fig. 28. Splitting of the input pulse in five secondary ones, exhibited by the numerical solution Eq. (7) with the coefficients shown in Eq. (54), and the initial condition (57). These coefficients correspond to the physical parameters given in Eqs. (47), and $E_{py} = 40 \ kV/cm$, $E_0 = -3.92098 \ kV/cm$ and $E_p = -2 \ kV/cm$.

The two examples presented in Figs. 27 and 28 show that the solutions of Eq. (7) corresponding to localized initial conditions may produce solutions which split into two, or more, solitary waves. This is not a surprise, as it is well-known that in the case of the integrable NLS equation, a sech input, depending on amplitude $B$ in front of it $[U(s, \zeta = 0) = B \ \text{sech} \ (bs)]$, may generate a soliton, or produce fission into a set of several solitary waves. Without any additional perturbation, the multi-soliton state would form a breather, while any perturbation splits it into a set of separating fundamental solitons. The number of solitons, $N$, is determined by what is usually called "area" $A$ of the input, *i.e.*, the integral of $U(s, \zeta = 0)$. This integral should not be confused with the *norm*, which is the integral of $[U(s, \zeta = 0)]^2$, and unlike the norm, the *area* is not a dynamical invariant of the NLS equation. In the case of a nonintegrable equation, such as Eq. (7), there is no exact criterion to predict the number of solitary waves generated by the splitting of the input. However, a qualitative similarity still holds: also in the case of Eq. (7) the number of fundamental solitons created by the splitting of a sech input increases as the *area A* of the input increases. We can observe that the value $A_{57} = 2.21$, corresponding to the input (57), is nearly twice as large in comparison to $A_{56} = 1.14$ for the pulse (56), and, as a consequence, the number of waves generated by the splitting increases from 2 to 5. It is interesting to observe that in the integrable case, if an initial pulse with *area A* splits into 2 solitary waves, a broader pulse with *area* $2A$ would split into $2 \times 2 = 4$ new waves. Therefore, the splitting of the pulse (57) into *five* new waves (instead of *four*), is an interesting manifestation of the nonintegrability of Eq. (7). On the other hand, an essential similarity with the integrable case, is that the initial pulse (57) splits into a set of solitary waves with *unequal amplitudes*. Therefore, the fission of solitary waves in the case of Eq. (7) presents some similarities to the results observed in the integrable cubic-NLS case, but the similarity is not exact due to the fact that Eq. (7) is not integrable.

## 5. Conclusion

The results presented in this communication show that solitons and breathers exist in biased photovoltaic pyroelectric photorefractive crystals, where the light propagation obeys Eq. (7). The application of the VA (variational approximation) predicts the existence of breathers with different shapes and oscillation periods. The VK (Vakhitov-Kolokolov) criterion correctly indicates that these solitons are stable solutions. The VA is elaborated on the basis of two different *ansätze*, which are defined in Eqs. (16) and (33). In both cases the averaged Lagrangians, given by Eqs. (18) and (34), respectively, involve the *dilogarithm* function $Li_2$. Therefore, we refer to these modes as *dilogarithmic breathers*. The existence of these VA-predicted breathers is then confirmed by full numerical solutions of Eq. (7). The numerical results demonstrate that, in addition to the regular (periodic) breathers, such as those shown in Figs. 18, 19 and 21, Eq. (7) also gives rise to robust breathers exhibiting beats (long-period modulations) in their intrinsic dynamics, such as the one shown in Fig. 24. The spectrum of oscillations of the beating breathers is nearly discrete, thus confirming that the oscillations are indeed subject to long-period quasi-periodic modulations. Another class of numerical solutions demonstrate spontaneous splitting of the input pulse into two or several secondary quasi-soliton pulses, as shown in Figs. 27 and 28. Systematic simulations of the splitting dynamics is a subject for a separate work, which is currently in progress.

### Declaration of competing interest

The authors declare that they have no known competing financial interests or personal relationships that could have appeared to influence the work reported in this paper.

### Credit authorship contribution statement

**C.A. Betancur-Silvera:** Investigation. Variational analysis. **A. Espinosa-Cerón:** Methodology. Numerical analysis. **B.A. Malomed:** Conceptualization. Formal analysis. Writing. **J. Fujioka:** Variational analysis. Organization. Writing.

### Acknowledgements

We thank DGTIC-UNAM (Dirección General de Cómputo y de Tecnologías de Información y Comunicación de la Universidad Nacional Autónoma de México) for granting us access to the computer *Miztli*, through Project LANCAD-UNAM-DGTIC-164, in order to carry out this work. Moreover, CABS acknowledges financial support from CONACYT (CVU 1184448). The work of BAM was supported, in part, by the Israel Science Foundation through grant No. 1695/22.